# Distributed Analysis and Load Balancing System for Grid Enabled Analysis on Hand-held devices using Multi-Agents Systems

Ahmad Naveed[4], Ali Arshad[4], Anjum Ashiq[4], Azim Tahir[4], Bunn Julian[1], Hassan Ali[4], Ikram Ahsan[4], Lingen Frank[1], McClatchey Richard[2], Newman Harvey[1], Steenberg Conrad[1], Thomas Michael[1], Willers Ian[3]

[1] California Institute of Technology (Caltech), Pasadena, CA 91125, USA
`{fvlingen,newman,conrad,thomas}@hep.caltech.edu,`
`Julian.Bunn@caltech.edu`
[2] University of the West of England, Bristol, UK
`Richard.mcclatchey@uwe.ac.uk`
[3] CERN
Geneva, Switzerland
`Ian.Willers@cern.ch`
[4] National University of Sciences and Technology, Rawalpindi, Pakistan
`{ahsan, ali.hassan, arshad.ali, ashiq.anjum, naveed.ahmad, tahir}@niit.edu.pk`

**Abstract.** Handheld devices, while growing rapidly, are inherently constrained and lack the capability of executing resource hungry applications. This paper presents the design and implementation of distributed analysis and load-balancing system for hand-held devices using multi-agents system. This system enables low resource mobile handheld devices to act as potential clients for Grid enabled applications and analysis environments. We propose a system, in which mobile agents will transport, schedule, execute and return results for heavy computational jobs submitted by handheld devices. Moreover, in this way, our system provides high throughput computing environment for hand-held devices.

## 1 Introduction

Handheld computing and wireless networks hold a great deal of promise in the fields of application development and ubiquitous data access. However, inherently constrained characteristics of mobile devices are the main hurdle.

Grids providing computational and storage resource sharing are the best implementation of distributed computing. We present here a brief description of a Grid enabled distributed analysis for handheld devices using mobile agents.

JASOnPDA and WiredOnPDA are the handheld physics analysis clients that we have developed and we will be using these two applications for analysis of our architecture.

## 2 Selection of Technologies

Grid enabled portal, JClarens [1], is used as a gateway to interact with Grid. For PocketPC we tried various JVMs and technologies, which included IBM Device Developer [5], Personal Profile, Super-Waba [6] and Savaje [7].

In the end we decided to go for PersonalJava and NSICOM's CrEme [8], as they suited our needs more than any other JVM.

For multi agent system we evaluated Aglets [3], DIET [4] and JADE-LEAP [2]. Due to small footprint, Personal Java compatibility and various other features we have chosen Jade-Leap for our research.

## 3 Architecture Design

Jade-Leap offers three types of containers that include Java based Jade-Leap J2SE containers, Personal Java based Jade-Leap PJava containers and MIDP based Jade-Leap MIDP containers.

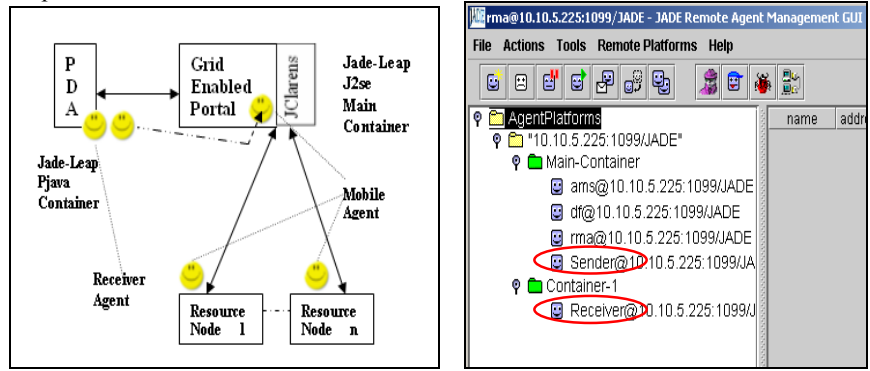

Figure 1 – (a) Architectural Overview          (b) JADE-LEAP Agent Management UI

As shown in Figure-1 (a), our architecture uses a combination of J2SE and PJava containers. Moreover, PJava containers running on handhelds must register with J2SE based Main-Container to communicate with other containers and platform, according to FIPA specifications.

After registering with the Main-Container, the client initiates two agents: the mobile (execution) agent and the receiver agent. While receiver agent resides on handheld device (Figure-1 (b)) waiting for response, mobile agents move to Main-Container (Figure-1 (b)) along with execution ontology and classes. Mobile agents execute job requests on server nodes or available resource nodes. Receiver Agent and Mobile Agent can intercommunicate to check job status, maintain connectivity, receiving results and killing jobs. After completing the job, the mobile agent either brings back the results or directly transfers them to handheld client.

### 3.1 Load Balancing

Every time an agent moves to a resource node for execution or storage it is provided with the load information of that node. From this information it deduces the load status. Mobile agents checks for the availability of required resources and also caters for the load of that machine. In case the load status notifies that resource node is over utilized the agent withdraws its execution or storage from that machine and looks for a under utilized node available in the farm of that node.

### 3.2 Fault Tolerance

While agents execute jobs remotely on behalf of handheld devices, they ensure reconnectivity by updating their location and state to their twin agent residing on handheld client.

On the server side architecture has been kept decentralized, there are more than one server nodes and handheld clients can register with all available servers. In case a sever goes down before job submission, handheld client can resubmit job to any other available node.

## 4 JASOnPDA / WiredOnPDA

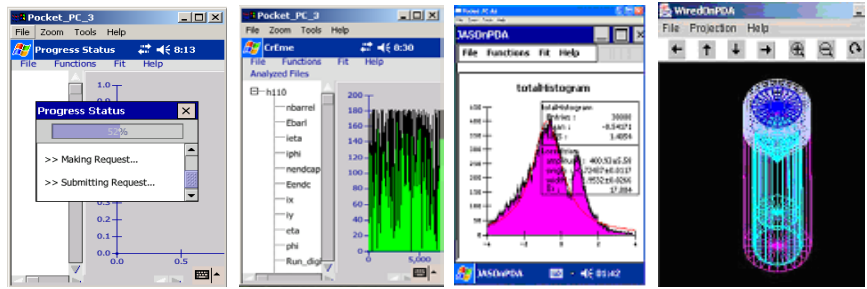

Figure 2 – JASOnPDA and WiredOnPDA Screen Shots

JASOnPDA (Figure 2) is a physics analysis tool used for analyzing data obtained from Linear Accelerator in the form of 1D-2D Histograms. Event data to be analyzed is in the form of ROOT files, which stores numerical physics data in a special, quickly accessible hierarchical structure.

WiredOnPDA (Figure 2) reads data from any source that either provides HepRep XML files or a HepRep-enabled HepEventServer. Files are parsed using a SAX XML parser. "Drawables" are then extracted from the parsed data and are displayed. This procedure is very resource and time consuming and thus makes it difficult to run on handheld devices.

## 5  Performance Analysis

We have carried out various statistical analysis using JASOnPDA and WiredOnPDA to derive conclusions about our proposed architecture.

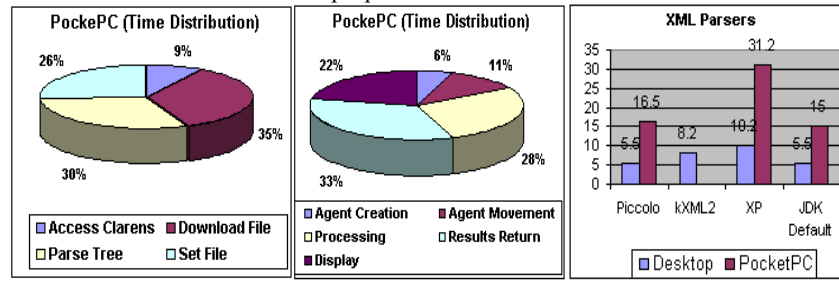

Figure 3 – Time distribution (a) without agents (b) with agents (c) XML Parsers Comparison

Figure 3(a) shows time taken by JASOnPDA when executed without mobile agents. Most of the time is taken in downloading large ROOT files to mobile device and parsing those files to retrieve numerical data. On the other hand in Figure 3(b) it can be seen that most of these time consuming tasks are now carried out at high power desktop machine, time consumption on the whole has also been reduced considerably.

Similarly, it was observed (Figure 3 (c)) that XML parsing (for WiredOnPDA) takes twice as much time on PocketPC as much on desktop. Therefore, using mobile agents we can carry out this task on desktop machines in less time.

## 6  Conclusions

This analysis environment has been developed to contribute to the scientists and physicist community to provide them access to data over wireless links. The slow performance of handheld devices was the major barrier that had to be over come to achieve true interactivity. This environment has solved these issues to a great extent for our existing applications.